\documentstyle[12pt,epsf]{article}
\def\lsim{\mathrel{\raise.3ex\hbox{$<$\kern-.75em\lower1ex\hbox{$\sim$}}}}
\def\gsim{\mathrel{\raise.3ex\hbox{$>$\kern-.75em\lower1ex\hbox{$\sim$}}}}
\newcommand{ \slashchar }[1]{\setbox0=\hbox{$#1$}   % set a box for #1
   \dimen0=\wd0                                     % and get its size
   \setbox1=\hbox{/} \dimen1=\wd1                   % get size of /
   \ifdim\dimen0>\dimen1                            % #1 is bigger
      \rlap{\hbox to \dimen0{\hfil/\hfil}}          % so center / in box
      #1                                            % and print #1
   \else                                            % / is bigger
      \rlap{\hbox to \dimen1{\hfil$#1$\hfil}}       % so center #1
      /                                             % and print /
   \fi}                                             %
\def\to{\rightarrow}

\def\be{\begin{equation}}
\def\ee{\end{equation}}
\def\bea{\begin{eqnarray}}
\def\eea{\end{eqnarray}}
\def\atversim#1#2{\lower0.7ex\vbox{\baselineskip\zatskip\lineskip\zatskip
  \lineskiplimit 0pt\ialign{$\matth#1\hfil##\hfil$\crcr#2\crcr\sim\crcr}}}

\renewcommand{\thefootnote}{\fnsymbol{footnote}}

\newcounter{appendixc}
\newcounter{subappendixc}[appendixc]
\newcounter{subsubappendixc}[subappendixc]

\renewcommand{\appendix}[1] {\vspace*{0.6cm}
        \refstepcounter{appendixc}
        \setcounter{figure}{0}
        \setcounter{table}{0}
        \setcounter{equation}{0}
        \renewcommand{\thefigure}{\Alph{appendixc}.\arabic{figure}}
        \renewcommand{\thetable}{\Alph{appendixc}.\arabic{table}}
        \renewcommand{\theappendixc}{\Alph{appendixc}}
        \renewcommand{\theequation}{\Alph{appendixc}.\arabic{equation}}
        \noindent{\bf Appendix \theappendixc #1}\par\vspace*{0.4cm}}

\begin{document}

\begin{titlepage}
\rightline{\vbox{\halign{&#\hfil\cr
%&hep-ph/xxxxxxx\cr
%&July 2000\cr
}}}
\begin{center}
{\Large\bf  Effects of anomalous couplings of the tau lepton}

\bigskip

\normalsize
{\bf  T. Huang, Z.-H. Lin and X. Zhang} \\
\vskip .3cm
Institute of High Energy Physics, Academia Sinica,\\
Beijing, 100039, P. R. China\\
\vskip .3in

\end{center}

\begin{abstract}
The dimension-six CP-conserving $SU_L(2)\times U_Y(1)$ invariant operators
involving the tau lepton are studied. It leads to the new physics effects
to the lepton flavor violation (LFV) and lepton flavor-changing neutral current
(FCNC). With the available experimental data on the decays $Z\rightarrow
\tau^+ + \tau^- $, $Z\rightarrow \mu^+ + \mu^- $ and $\mu^- \rightarrow e^-
+\gamma$,
the constraints on parameters of the effective lagrangian can be given.
According to such constraints, the branching ratio of the decay
$Z\rightarrow \mu^\pm + \tau^\mp $
could reach to $10^{-6}-10^{-7}$.
%the corrections to lepton
%anomalous magnetic moments and the effects on the processes
%$\tau^-\rightarrow\mu^-+\gamma$, $\tau^-\rightarrow e^-+\gamma$
%are much smaller than the present experimental data.
The size of the LFV effects depends crucially on the
dynamics of the lepton mass generation. Assuming the lepton mass
matrices in the form of a Fritzsch ansatz, we point out that the experiment on
$\mu\to e \gamma$ will put stronger limits on the anomalous magnitic and electric dipole moments
of the tau lepton than obtained by  Escribano and Mass\'{o}.
\end{abstract}

\renewcommand{\thefootnote}{\arabic{footnote}}
\end{titlepage}
%%%%%%%%%%%%%%%%%%%%%%%%%%%%%%%%%%%%%%%%%p1%%%%%%%%%%%%%%%%%%%%%%%%%%%%%%%%%

\section {Introduction}

Although the Standard Model (SM) has been successful in describing
the physics of the
electroweak interaction~\cite{Hewett}, it is quite possible that the SM is only
an effective theory
which breaks down at higher energies as the deeper structure of the underlying
physics emerges.

There are reasons to believe that the deviation from the SM might
first appear in the interactions involving the third-generation
fermions~\cite{Whisnant,Yang,Zhang}. The tau
lepton is possibly special probe of new physics in the lepton sector due to the
fact that it is the only lepton which is heavy enough to have
hadronic decays and
that the heavier fermions are more sensitive to the new physics related
to mass generation.

There has been extensive studies on the
tau physics and related possible new physics of various extension of the SM models.
We take a model-independent approach and formulate new physics effects in terms of
an effective lagrangian.

%%%%%%%%%%%%%%%%%%%%%%%%%%%%%%%%%%%%%%%%%p2%%%%%%%%%%%%%%%%%%%%%%%%%%%%%%%%%
Without specifying the detail of the underlying new physics,
the effective lagrangian to dimension 6 can be written as ,
\begin{eqnarray}\label{eq1}
{\cal L}_{eff}={\cal L}_0+\frac{1}{\Lambda^2}\sum_i C_i O_i,
\end{eqnarray}
where ${\cal L}_0$ is the SM Lagrangian, $\Lambda$ is the new physics
scale and $O_i$ involve only $( \nu_{\tau}, ~ \tau_L ), ~ \tau_R$,
the gauge and the scalar bosons and are
$SU_c(3)\times SU_L(2)\times U_Y(1)$ invariant.
$C_i$ are
constants which represent the coupling strengths of $O_i$~\cite{Burgess}.
A complete list of CP-violating operators of leptons
has been given in Ref.~\cite{Young}.

To restrict ourselves to the lowest order, we consider only tree diagrams
and to the order of $1/\Lambda^2$, so only one
vertex in a given diagram can contain anomalous couplings. Under these
conditions, operators connected by the field equations
are not independent. The fermion
and the Higgs boson equations of motion
can be used to list the operators but the equations of
motion of the gauge bosons can not.
Furthermore, all the operators $O_i$ are Hermitian and the
coefficients $C_i$ are real, since we assume that
the available energies are below the unitarity cuts of new-physics
particles and no imaginary part can be generated by the new physics
effect.
%%%%%%%%%%%%%%%%%%%%%%%%%%%%%%%%%%%%%%%%%p3%%%%%%%%%%%%%%%%%%%%%%%%%%%%%%%%%
The higher dimension operators modify the couplings of the tau lepton and
gauge (or Higgs) bosons, and generate also lepton flavor violation interaction.
However, the size of the lepton flavor violating effect depends crucially on the
lepton mass mixing matrices.
We take the lepton mass ansatz of Fritzsch, which was discussed
twenty years ago~\cite{Fritzsch} and developed in lepton sector
in recent years~\cite{Xing1,Xing2}. This ansatz is based on "democratic symmetry" of
 fermion mass matrix, which has the form
\begin{eqnarray}
M_{0i}=c_{0i}
\left (
\begin{array}{lcr}
1 & 1 & 1\\
1 & 1 & 1\\
1 & 1 & 1
\end{array}
\right ).
\end{eqnarray}
Here $i$ stands for $ u, d$ in case of quarks and $l$ in case
of the charged leptons. For the charged leptons, this matrix could be
diagonalized as
\begin{eqnarray}
M_H^l=c_l
\left (
\begin{array}{lcr}
0 & 0 & 0\\
0 & 0 & 0\\
0 & 0 & 3
\end{array}
\right )
\end{eqnarray}
implying that only the third family lepton has nonzero mass $3c_l$
under the democratic limit.

%%%%%%%%%%%%%%%%%%%%%%%%%%%%%%%%%%%%%%%%%p4%%%%%%%%%%%%%%%%%%%%%%%%%%%%%%%%%
We study the correlated effects of anomalous couplings of the tau lepton, such as
%$Z {\bar \tau} \tau$ and $Z {\bar \tau} \mu$. we will point out that
%with the experimental
%data at LEP on the decays  $Z\rightarrow \tau^+ + \tau^- $, $Z\rightarrow
%\mu^+ + \mu^- $,
% the branching ratio of the decay $Z\rightarrow
% \mu^\pm + \tau^\mp $ could be as large as
% $10^{-6}-10^{-7}$, which is experimentally accessible in the future
% "Z-Factory".
\begin{itemize}
\item Anomalous magnetic and electric dipole moments of the tau lepton by allowing
the mixing of three generations in the lepton sector;
\item The decay $Z\rightarrow \mu^\pm + \tau^\mp $ with the experimental data at LEP
on the decays $Z\rightarrow \tau^+ + \tau^- $, $Z\rightarrow\mu^+ + \mu^- $;
\item Others.
\end{itemize}

%%%%%%%%%%%%%%%%%%%%%%%%%%%%%%%%%%%%%%%%%p5%%%%%%%%%%%%%%%%%%%%%%%%%%%%%%%%%
%%%%%%%%%%%%%%%%%%%%%%%%%%%%%%%%%%%%%%%%%p6%%%%%%%%%%%%%%%%%%%%%%%%%%%%%%%%%
\section {Anomalous couplings of the tau lepton}
The expressions of the CP-conserving operators involving the third family
leptons are parallel to their corresponding ones involving the third
family quarks, but
the number of independent operators is much less due to the absence of
right-handed neutrino and the strong interactions.
%The standard notation is:
%$L$ denotes the third family left-handed doublet leptons, $\Phi$ is
%the Higgs doublet, $W_{\mu\nu}$ and $B_{\mu\nu}$ are the SU(2) and
%U(1) gauge boson field tensors in the appropriate matrix forms, and
%$D_\mu$ denotes the appropriate covariant derivatives.
The possible CP-conserving $SU_L(2)\times U_Y(1)$
invariant operators are given by

\begin{eqnarray}
\label{operator}
O_{LW}&=&\left [\bar L \gamma^{\mu}\tau^I D^{\nu}L
         +\overline{D^{\nu}L} \gamma^{\mu}\tau^I L\right ] W^I_{\mu\nu},\nonumber\\
O_{LB}&=&\left [\bar L \gamma^{\mu} D^{\nu}L
         +\overline{D^{\nu}L} \gamma^{\mu} L\right ] B_{\mu\nu},\nonumber\\
O_{\tau B}&=&\left [\bar \tau_R \gamma^{\mu} D^{\nu}\tau_R
         +\overline{D^{\nu}\tau_R} \gamma^{\mu}\tau_R\right ]
         B_{\mu\nu},\nonumber\\
O_{\Phi L}^{(1)}&=&i\left [\Phi^{\dagger}D_{\mu}\Phi
                   -(D_{\mu}\Phi)^{\dagger}\Phi\right ] \bar L
                  \gamma^{\mu}L,\nonumber\\
O_{\Phi L}^{(3)}&=&i\left [\Phi^{\dagger}\tau^I D_{\mu}\Phi
                   -(D_{\mu}\Phi)^{\dagger}\tau^I\Phi\right ]
                   \bar L \gamma^{\mu}\tau^I L,\nonumber\\
O_{\Phi \tau}&=&i\left [\Phi^{\dagger}D_{\mu}\Phi
             -(D_{\mu}\Phi)^{\dagger}\Phi\right ]
             \bar \tau_R \gamma^{\mu}\tau_R,\nonumber\\
O_{\tau 1}&=&(\Phi^{\dagger}\Phi-\frac{v^2}{2})\left [\bar L \tau_R\Phi
         +\Phi^{\dagger}\bar \tau_R L\right ],\nonumber\\
O_{D\tau}&=&\left [(\bar L D_{\mu} \tau_R) D^{\mu}\Phi
         +(D^{\mu}\Phi)^{\dagger}(\overline{D_{\mu}\tau_R}L)\right ],\nonumber\\
O_{\tau W\Phi}&=&\left [(\bar L \sigma^{\mu\nu}\tau^I \tau_R) \Phi
         +\Phi^{\dagger}(\bar \tau_R \sigma^{\mu\nu}\tau^I L)\right ]
          W^I_{\mu\nu},\nonumber\\
O_{\tau B\Phi}&=&\left [(\bar L \sigma^{\mu\nu} \tau_R) \Phi
         +\Phi^{\dagger}(\bar \tau_R \sigma^{\mu\nu} L)\right ]
          B_{\mu\nu}.
\end{eqnarray}
The standard notation is:\\
\begin{tabular}{ll}
%   &  \\
$L$       & the third family left-handed doublet leptons,\\
$\Phi$       & the Higgs doublet,\\
$W_{\mu\nu}$ & the SU(2)gauge boson field tensors,\\
$B_{\mu\nu}$ & the U(1) gauge boson field tensors ,\\
$D_\mu$      & the appropriate covariant derivatives.
\end{tabular}

%%%%%%%%%%%%%%%%%%%%%%%%%%%%%%%%%%%%%%%%%p7%%%%%%%%%%%%%%%%%%%%%%%%%%%%%%%%%
The expressions of these  operators
after electroweak symmetry breaking  in the unitary gauge
are given by
\begin{eqnarray}
O_{LW}&=&\frac{1}{2}W^3_{\mu\nu}
\left [\bar \nu_{\tau}\gamma^{\mu}P_L\partial^{\nu}\nu_{\tau}
+\partial^{\nu}\bar \nu_{\tau}\gamma^{\mu}P_L\nu_{\tau}
-\bar \tau \gamma^{\mu}P_L \partial^{\nu}\tau
-\partial^{\nu}\bar \tau\gamma^{\mu}P_L\tau\right ]
  \nonumber\\
&& +\frac{1}{\sqrt 2}
\left [W^+_{\mu\nu}(\bar \nu_{\tau}\gamma^{\mu}P_L\partial^{\nu}\tau
+\partial^{\nu}\bar \nu_{\tau}\gamma^{\mu}P_L\tau)
+W^-_{\mu\nu}(\bar \tau\gamma^{\mu}P_L\partial^{\nu}\nu_{\tau}
+\partial^{\nu}\bar \tau\gamma^{\mu}P_L\nu_{\tau})\right ]\nonumber\\
& & -ig_2\bar L \gamma^{\mu} \left [W_{\mu},W_{\nu}\right ]\partial^{\nu}L
  -ig_2\partial^{\nu}\bar L \gamma^{\mu}
  \left [W_{\mu},W_{\nu}\right ]L
%\nonumber\\
 -ig_2\bar L \gamma^{\mu} \left [W_{\mu\nu},W^{\nu}\right ] L,\nonumber\\
O_{LB}&=&B_{\mu\nu}\left [\bar L\gamma^{\mu} \partial^{\nu}L
   +\partial^{\nu} \bar L\gamma^{\mu}L\right ],\nonumber\\
O_{\tau B}&=&\left [\bar \tau\gamma^{\mu}P_R \partial^{\nu}\tau
         +\partial^{\nu} \bar \tau\gamma^{\mu}P_R\tau\right ]B_{\mu\nu},\nonumber\\
O_{\Phi L}^{(1)}&=&\frac{m_Z}{v}(H+v)^2 Z_{\mu}
        \left [\bar \nu_{\tau}\gamma^{\mu}P_L\nu_{\tau}
          +\bar \tau\gamma^{\mu}P_L\tau\right ],\nonumber \\
O_{\Phi L}^{(3)}&=&-\frac{m_Z}{v}(H+v)^2 Z_{\mu}
         \left [\bar\nu_{\tau}\gamma^{\mu}P_L\nu_{\tau}
          -\bar \tau\gamma^{\mu}P_L\tau\right ]           \nonumber\\
& &      +\frac{1}{\sqrt 2}g_2(H+v)^2
        \left [W^+_{\mu}\bar\nu_{\tau}\gamma^{\mu}P_L\tau
        +W^-_{\mu}\bar\tau\gamma^{\mu}P_L\nu_{\tau}\right ],\nonumber\\
O_{\Phi \tau}&=&\frac{m_Z}{v}(H+v)^2 Z_{\mu}\bar\tau\gamma^{\mu}P_R\tau,\nonumber\\
O_{\tau 1}&=&\frac{1}{2\sqrt 2}H(H+v)(H+2v)\bar \tau \tau,\nonumber\\
O_{D\tau}&=&\frac{1}{2\sqrt 2}\partial^{\mu}H \left [
   \partial_{\mu}(\bar \tau\tau)
   +\bar \tau\gamma_5\partial_{\mu}\tau
   -(\partial_{\mu}\bar \tau)\gamma_5\tau
   +2g_1B_{\mu}\bar \tau i\gamma_5\tau\right ]\nonumber\\
& & +\frac{i}{2\sqrt 2}\frac{m_Z}{v} (H+v)Z^{\mu}
   \left [(\partial_{\mu}\bar \tau)\tau-\bar\tau\partial_{\mu}\tau
   -\partial_{\mu}(\bar \tau\gamma_5 \tau)
   -i2g_1B_{\mu}\bar \tau\tau\right ]\nonumber\\
& & -\frac{i}{2}g_2(H+v)
\left [W_{\mu}^+ (\bar \nu_{\tau}P_R \partial^{\mu}\tau
    +ig_1 B^{\mu} \bar \nu_{\tau}P_R  \tau) \right.      \nonumber\\
& & \left. -W_{\mu}^- ( \partial^{\mu}\bar \tau P_L \nu_{\tau}-
    ig_1 B^{\mu} \bar \tau P_L \nu_{\tau})\right ],\nonumber\\
O_{\tau W\Phi}&=&\frac{1}{2}(H+v)
\left [W^+_{\mu\nu}(\bar \nu_{\tau} \sigma^{\mu\nu}P_R \tau)
    +W^-_{\mu\nu}(\bar \tau\sigma^{\mu\nu}P_L \nu_{\tau})
    -\frac{1}{\sqrt 2}W^3_{\mu\nu}
    (\bar \tau \sigma^{\mu\nu} \tau)\right.\nonumber\\
& & +ig_2(W^+_{\mu}W^3_{\nu}-W^3_{\mu}W^+_{\nu})
(\bar \nu_{\tau} \sigma^{\mu\nu} P_R \tau)
    -ig_2(W^-_{\mu}W^3_{\nu}-W^3_{\mu}W^-_{\nu})
(\bar \tau \sigma^{\mu\nu}P_L \nu_{\tau}) \nonumber\\
& &\left. -i\frac{g_2}{\sqrt 2} (W^+_{\mu}W^-_{\nu}-W^-_{\mu}W^+_{\nu})
      (\bar \tau \sigma^{\mu\nu}\tau)\right ],\nonumber\\
O_{\tau B\Phi}&=&\frac{1}{\sqrt 2}(H+v) B_{\mu\nu}
            (\bar \tau \sigma^{\mu\nu}\tau).
\end{eqnarray}

%%%%%%%%%%%%%%%%%%%%%%%%%%%%%%%%%%%%%%%%%p8%%%%%%%%%%%%%%%%%%%%%%%%%%%%%%%%%
The possibilities of contributions of the dimension-six CP-conserving
operators
to some three-particle couplings are shown in Table 1.

\begin{table}[thb]
\begin{center}
\begin{tabular}{|l|c|c|c|c|}
\hline
 & & & & \\
 &$~W\nu \tau ~$ & $~Z\tau \tau~$ & $~\gamma \tau\tau~$ & $~H\tau\tau~$ \\
 & & & & \\ \hline

  $~~~O_{\Phi L}^{ (3)}$ &$\times$&$\times$&        &           \\ \hline
  $~~~O_{LW}$            &$\times$&$\times$&$\times$&           \\ \hline
  $~~~O_{D\tau}$         &$\times$&$\times$&        &$\times$   \\ \hline
  $~~~O_{\tau W\Phi}$    &$\times$&$\times$&$\times$&           \\ \hline
  $~~~O_{LB}$            &        &$\times$&$\times$&           \\ \hline
  $~~~O_{\tau B}$        &        &$\times$&$\times$&           \\ \hline
  $~~~O_{\tau B\Phi}$    &        &$\times$&$\times$&           \\ \hline
  $~~~O_{\Phi L}^{ (1)}$ &        &$\times$&        &           \\ \hline
  $~~~O_{\Phi \tau}$     &        &$\times$&        &           \\ \hline
  $~~~O_{\tau 1}$        &        &        &        &$\times$   \\ \hline
\end{tabular}
\end{center}
\caption[]{
The contribution status of dimension-six CP-conserving operators
to the tau couplings.
The contribution of a CP-conserving operator to a particular vertex is
marked by $\times$.
}
\end{table}

%%%%%%%%%%%%%%%%%%%%%%%%%%%%%%%%%%%%%%%%%p9%%%%%%%%%%%%%%%%%%%%%%%%%%%%%%%%%
According to their contribution to the three-particle vertices of
charged and neutral current, we classify the operators as: \\
 Class A:  $O_{\Phi L}^{ (3)}$,
           $O_{LW}$,
           $O_{D\tau}$
           and $O_{\tau W\Phi}$,
           contributing to both charged and neutral currents.\\
 Class B:  $O_{\Phi L}^{ (1)}$,
           $O_{\Phi \tau}$,
           $O_{LB}$,
           $O_{\tau B}$
           and $O_{\tau B\Phi}$,
           contributing to neutral currents.\\
 Class C:  $O_{\tau 1}$,
           no contribution to charged and neutral currents.

Since Class C operators contribute only to the $H\tau\tau$ coupling, they
may not be probed at future colliders. We do not consider
these operators here.
Both Class A and Class B operators affect neutral currents of the tau.
%associated with the decay $Z\rightarrow\tau^+ + \tau^- $ which we will
%discuss in the following sections.
%%%%%%%%%%%%%%%%%%%%%%%%%%%%%%%%%%%%%%%%%p9%%%%%%%%%%%%%%%%%%%%%%%%%%%%%%%%%

Collecting all the relevant terms we get the effective CP-conserving
couplings,

\begin{eqnarray}
{\cal L}_{W\nu\tau}&=&\frac{C_{\Phi L}^{(3)}}{\Lambda^2}\frac {g_2}
        {\sqrt2}v^2 W^+_{\mu}(\bar \nu_{\tau} \gamma^{\mu}P_L \tau)
         -\frac{C_{D\tau}}{\Lambda^2}\frac{v}{\sqrt 2}\frac{g_2}{\sqrt 2}
               W_{\mu}^+ (i\bar \nu_{\tau} P_R \partial^{\mu}\tau)\nonumber\\
           & &+\frac{C_{\tau W\Phi}}{\Lambda^2}\frac{v}{2}
                 W^+_{\mu\nu}(\bar \nu_{\tau} \sigma^{\mu\nu}P_R \tau)
                  +\frac{C_{LW}}{\Lambda^2} \frac{1}{\sqrt 2}
                          W^+_{\mu\nu}(\bar \nu_{\tau}\gamma^{\mu}P_L\partial^{\nu}\tau
                                +\partial^{\nu}\bar \nu_{\tau}\gamma^{\mu}P_L \tau),\nonumber\\
%----------------------------------------------------------------------
{\cal L}_{Z\tau\tau}&=&(\frac{C_{\Phi L}^{(1)}}{\Lambda^2}
  +\frac{C_{\Phi L}^{(3)}}{\Lambda^2})(v m_Z)
      Z_{\mu}(\bar \tau\gamma^{\mu}P_L \tau)
        +\frac{C_{\Phi \tau}}{\Lambda^2}(v m_Z)
        Z_{\mu}(\bar \tau \gamma^{\mu}P_R \tau)\nonumber\\
        & & +(\frac{C_{LW}}{\Lambda^2}\frac{c_W}{2}
                          +\frac{C_{LB}}{\Lambda^2}s_W)
                     Z_{\mu\nu}(\bar \tau \gamma^{\mu}P_L \partial^{\nu}\tau
                                       +\partial^{\nu}\bar \tau \gamma^{\mu}P_L \tau)\nonumber\\
                               & &+\frac{C_{\tau B}}{\Lambda^2}s_W
                                   Z_{\mu\nu}(\bar \tau \gamma^{\mu}P_R \partial^{\nu}\tau
                                               +\partial^{\nu} \bar \tau\gamma^{\mu}P_R \tau)\nonumber\\
                                       & &+\frac{m_Z}{2\sqrt 2}
                                          \frac{C_{D\tau}}{\Lambda^2}
                                            Z^{\mu}\left [i(\partial_{\mu}\bar \tau
                                          \tau-\bar \tau\partial_{\mu}\tau)
                                                      -i\partial_{\mu}(\bar \tau \gamma_5 \tau) \right ]\nonumber\\
                                                  & &+(\frac{C_{\tau W\Phi}}{\Lambda^2}\frac{c_W}{2}
                                                  +\frac{C_{\tau B\Phi}}{\Lambda^2}s_W)\frac{v}{\sqrt 2}
                                                     Z_{\mu\nu}(\bar \tau \sigma^{\mu\nu} \tau),\nonumber\\
%----------------------------------------------------------------------------
{\cal L}_{\gamma \tau\tau}&=&
  (\frac{C_{LB}}{\Lambda^2}c_W-\frac{C_{LW}}{\Lambda^2}\frac{s_W}{2})
    A_{\mu\nu}(\bar \tau \gamma^{\mu}P_L \partial^{\nu}\tau
               +\partial^{\nu}\bar \tau \gamma^{\mu}P_L \tau)\nonumber\\
           & &+\frac{C_{\tau B}}{\Lambda^2}c_W
              A_{\mu\nu}(\bar \tau \gamma^{\mu}P_R \partial^{\nu}\tau
                   +\partial^{\nu} \bar \tau\gamma^{\mu}P_R \tau)\nonumber\\
               & &+(\frac{C_{\tau B\Phi}}{\Lambda^2}c_W
                    -\frac{C_{\tau W\Phi}}{\Lambda^2}\frac{s_W}{2})\frac{v}{\sqrt 2}
                   A_{\mu\nu}(\bar \tau \sigma^{\mu\nu} \tau),
                   \end{eqnarray}
                   where $P_{L,R}\equiv(1\mp \gamma_5)/2$,
                   $Z_{\mu}=-\cos\theta_W W^3_{\mu}+\sin\theta_W B_{\mu}$,
                   $A_{\mu}=\sin\theta_W W^3_{\mu}+\cos\theta_W B_{\mu}$,
                   $s_W\equiv \sin\theta_W$ and $c_W\equiv\cos\theta_W$.
%=======================================================================
%%%%%%%%%%%%%%%%%%%%%%%%%%%%%%%%%%%%%%%%%p10%%%%%%%%%%%%%%%%%%%%%%%%%%%%%%%%%
Now we can write down the $Z\tau\tau$ and $\gamma\tau\tau$ vertices
including both the SM couplings and new physics effects as
%-------------------------------------------------------------------------
\begin{eqnarray}\label{ver}
 \Gamma_{\mu}^{Z,\gamma}=
 -ieg^{Z,\gamma}
 \left[\gamma_{\mu}V^{Z,\gamma}
 -\gamma_{\mu}\gamma_5  A^{Z,\gamma}
 -\frac{1}{2m_\tau}(i k_{\nu}\sigma^{\mu\nu}) S^{Z,\gamma} \right ],
 \end{eqnarray}
 %----------------------------------------------------------------------
 where $ g^{Z}=1/(4s_Wc_W)$, $g^{\gamma}=1$,
 and $k=p_{\tau}+p_{\bar \tau}$ ($p_\tau$ and $p_{\bar
 \tau}$ are the
 momenta of outgoing tau and anti-tau respectively).
 We neglect the scalar and pseudoscalar couplings, $k_{\mu}$ and
 $k_{\mu}\gamma_5$, since
  these terms give contributions proportional
  to the electron mass in $e^+e^-$ collisions.
  The vector and axial-vector couplings $V^{Z,\gamma}$ and $A^{Z}$
  contain both the SM and new physics contributions, while $A^{\gamma}$
 and $S^{Z,\gamma}$ contain only new physics contributions.
 One can write the vector and axial-vector couplings as
 %-----------------------------------------------------------------
 \begin{eqnarray}
 V^{Z,\gamma}=(V^{Z,\gamma})^0+\delta V^{Z,\gamma},~~~~~
 A^{Z,\gamma}=(A^{Z,\gamma})^0+\delta A^{Z,\gamma},
 \end{eqnarray}
 %---------------------------------------------------------------------
 where $(V^{Z,\gamma})^0$ and $(A^{Z,\gamma})^0$ represent the SM
 couplings
 and $\delta V^{Z,\gamma},\delta A^{Z,\gamma}$ the anomalous new
 physics contributions.
 At the tree level the SM couplings are
 $(V^{Z,\gamma})^0=1-4s_W^2,1$ and
 $(A^{Z,\gamma})^0=1,0$,
 while the new physics contributions are given by
 %-------------------------------------------------------------------
 %-------------------------------------------------------------------
 \begin{eqnarray}
 \delta V^Z&=&\frac{2s_Wc_W}{e}\frac{vm_Z}{\Lambda^2}\left [
    C_{LW}\frac{c_Wk^2}{2vm_Z}
       +(C_{LB}+C_{\tau B})\frac{s_Wk^2}{vm_Z}
          -C^{(1)}_{\Phi L}-C^{(3)}_{\Phi L}-C_{\Phi \tau} \right ],\nonumber \\
      %---------------------------------------------------------------------
      \delta A^Z&=&\frac{2s_Wc_W}{e}\frac{vm_Z}{\Lambda^2}\left [
         C_{LW}\frac{c_Wk^2}{2vm_Z}
       +(C_{LB}-C_{\tau B})\frac{s_Wk^2}{vm_Z}
         -C^{(1)}_{\Phi L}-C^{(3)}_{\Phi L}+C_{\Phi \tau}\right ],\nonumber \\
     %---------------------------------------------------------------------
         \label{eq32}
  S^Z&=&-\frac{8s_Wc_W}{e}\frac{m_\tau}{\Lambda^2}\frac{v}{\sqrt 2}\left [
             C_{D\tau}\frac{m_Z}{2v}
        -C_{\tau W\Phi}c_W-2C_{\tau B\Phi}s_W\right ],\nonumber\\
    %----------------------------------------------------------------------
            \label{eq13}
    \delta V^{\gamma}&=&\frac{1}{e}\frac{k^2}{2\Lambda^2}\left [
               C_{LW}\frac{s_W}{2}-(C_{LB}+C_{\tau B})c_W
              \right ],\nonumber\\
\delta A^{\gamma}&=&\frac{1}{e}\frac{k^2}{2\Lambda^2}\left [
   C_{LW}\frac{s_W}{2}-(C_{LB}-C_{\tau B})c_W
   \right ],\nonumber \\
\label{eq15}
 S^{\gamma}&=&\frac{2m_\tau}{e}\frac{\sqrt 2 v}{\Lambda^2}\left [
    C_{\tau W\Phi}\frac{s_W}{2}-C_{\tau B\Phi}c_W \right ].
    \end{eqnarray}
    %--------------------------------------------------------------------------
%%%%%%%%%%%%%%%%%%%%%%%%%%%%%%%%%%%%%%%%%p11%%%%%%%%%%%%%%%%%%%%%%%%%%%%%%%%%

Now we discuss the neutral current processes involving three-family
charged leptons. With assumption that the new
physics correction only resides in the third-family lepton,
the neutral currents effective lagrangian has a non-universal form in
three-family lepton sector, like
%-------------------------------------------------------------------
\begin{eqnarray}\label{eq36}
{\cal L}_{eff}=&&
\frac{2m_Z}{v} Z^\mu
{{ \pmatrix{{\overline e}
  \cr {\overline \mu} \cr
    {\overline
       \tau} \cr}}}^T
        U_L^{(l)}
    \left \{
    g_L
    \pmatrix{1& &\cr
    & 1 & \cr
    &   & 1 \cr}
 + \pmatrix{ 0 & & \cr
      & 0 & \cr
        & &  \delta_L \cr }
      \right \}
     U_L^{ \dagger (l)}
     \gamma_\mu
     P_L\pmatrix{e \cr \mu \cr \tau \cr}
\nonumber \\
&& +(L\longleftrightarrow R) ,
 \end{eqnarray}
 %------------------------------------------------------------------------
 where L (R) denotes the left (right) hand, the SM tree level couplings
 are
 \begin{eqnarray}
 g_L&=&-{1 \over 2} +\sin^2\theta_W,\nonumber\\
 g_R&=&\sin^2 \theta_W,
 \end{eqnarray}
 %----------------------------------------------------------------------
 the new physics couplings are
 \begin{eqnarray}
 \delta_L&=&{v^2 \over 2\Lambda^2 }
         (C^{(1)}_{\Phi L}+C^{(3)}_{\Phi L}
         -\frac{C_{LB} s_W k^2}{v m_Z}
         -\frac{C_{L W}c_W k^2}{2v m_Z}),\nonumber\\
     \delta_R&=&{v^2 \over 2\Lambda^2 }
         (C_{\Phi \tau}
         -\frac{C_{\tau B}s_W k^2}{v m_Z}),
 \end{eqnarray}
%------------------------------------------------------------------------
and $U_{L,R}^{(l)}$ are unitary flavor-mixing matrices diagonalizing
the left-handed and right-handed charged leptons respectively.

Here, we
classify the effective lagrangian not as vector part and axial-vector part
but as left hand part and right hand part by the reason that $\delta_L$
and
$\delta_R$ are independent by containing different parameters.
Noticing that the SM corresponds to  ${\cal L}^{eff}$ in the limit
$\Lambda \rightarrow \infty$, the corrections of new physics vanish and
$U_{L,R}^{(l)}$ are not measurable in neutral current processes in the SM.

%%%%%%%%%%%%%%%%%%%%%%%%%%%%%%%%%%%%%%%%%p12%%%%%%%%%%%%%%%%%%%%%%%%%%%%%%%%%
\section {Fritzsch Ansatz on the lepton mass}

The flavor-mixing matrix
$U$ (=$U^{(l)}_L = U^{(l)}_R$ ) which diagonalizes the "democratic matrix"
$M_{0l}$ by the transformation $U M_{0l} U^{\dagger}= M_H^l$, is given by~\cite{Xing3}
%------------------------------------------------------------------
\begin{eqnarray}
U \; =\; \left ( \begin{array}{ccc}
\displaystyle\frac{1}{\sqrt{2}} & -\displaystyle\frac{1}{\sqrt{2}}      &
0 \nonumber\\
\displaystyle\frac{1}{\sqrt{6}} & \displaystyle\frac{1}{\sqrt{6}}       &
-\displaystyle\frac{2}{\sqrt{6}} \nonumber\\
\displaystyle\frac{1}{\sqrt{3}} & \displaystyle\frac{1}{\sqrt{3}}       &
\displaystyle\frac{1}{\sqrt{3}}
\end{array} \right ) \; .
\end{eqnarray}
%------------------------------------------------------------------------
Then we have
\begin{eqnarray}\label{eq42}
U\left (
\begin{array}{lcr}
0 &  & \\
& 0 & \\
 &  & {\delta_L}
 \end{array}
 \right )U^{\dagger}
 =\left (
 \begin{array}{lcr}
 0&0&0\\
 0&\frac{2}{3}\delta_L&-\frac{\sqrt 2}{3}\delta_L\\
 0&-\frac{\sqrt2}{3}\delta_L&\frac{1}{3}\delta_L
 \end{array}
 \right ).
 \end{eqnarray}
 %---------------------------------------------------------------------
 %---------------------------------------------------------------------
 In this approximation, there is no extra correction to the vertices of
 $Zee$, $Ze\mu$ and $Ze\tau$. However, some small corrections proportional
 to $\delta_L$ are generated naturally in the second and third families.
%%%%%%%%%%%%%%%%%%%%%%%%%%%%%%%%%%%%%%%%%p13%%%%%%%%%%%%%%%%%%%%%%%%%%%%%%%%%

By taking into account the existence of the mass of electron and muon, the
democratic symmetry would be broken and there is
\begin{eqnarray}
M_{l} \; =\; c_{l} \left ( \begin{array}{ccc}
1       & 1     & 1 \\
1       & 1     & 1 \\
1       & 1     & 1
\end{array} \right ) + \Delta M_{l} ,
\end{eqnarray}
%-----------------------------------------------------------------------
where $\Delta M_{l}$ are the symmetry breaking terms for the charged
leptons. $\Delta M_{l}$ would be a diagonal mass shift, i.e.
%---------------------------------------------------------------------
%{\large
\begin{eqnarray}
\Delta M_{l} \; =\; \left ( \begin{array}{ccc}
\delta_{l}      & 0     & 0 \\
0       & \varrho^{~}_{l}       & 0 \\
0       & 0     & \varepsilon^{~}_{l}
\end{array} \right ) ,
\end{eqnarray}
%}
and the neutrino mass matrix $M_{\nu}$ has the diagonal form with three
different eigenvalues.
%----------------------------------------------------------------------
Following Ref.~\cite{Xing1}, we take $\delta_{l}=-\varrho_l$ to
simplify the problem, and, the mixing angles can be completely
expressed in terms of the charged lepton mass. The flavor-mixing  matrix
$V$
($V M_l V^{\dagger}=M_{diag}$) can be parametrized in terms of
three Euler
angles as follow
%--------------------------------------------------------------------
\begin{eqnarray}
V \; =\; \left ( \begin{array}{ccc}
c_{12}c_{13}                            & s_{12}c_{13}
& s_{13} \\
-s_{12}c_{23}-c_{12}s_{23}s_{13}        & c_{12}c_{23}-s_{12}s_{23}s_{13}
& s_{23}c_{13} \\
s_{12}s_{23}-c_{12}c_{23}s_{13}         & -c_{12}s_{23}-s_{12}c_{23}s_{13}
&c_{23}c_{13}
\end{array} \right ) ,
\end{eqnarray}
%--------------------------------------------------------------------------
where $s_{ij}\equiv \sin\theta_{ij}$, $c_{ij}\equiv \cos\theta_{ij}$ and
the three mixing angles are determined as follows
%---------------------------------------------------------------------------
\begin{eqnarray}
\tan\theta_{12}&=&-1+\frac{2}{\sqrt{3}}\sqrt{\frac{m_{e}}{m_{\mu}}} ,\nonumber\\
\tan\theta_{23}&=&-\sqrt{2}-\frac{3}{\sqrt{2}}\frac{m_{\mu}}{m_{\tau}},\nonumber\\
\tan\theta_{13}&=&-\frac{2}{\sqrt{6}}\sqrt{\frac{m_{e}}{m_{\mu}}}.\nonumber
\end{eqnarray}
%-----------------------------------------------------------------------
Then we have
%{\large
\begin{eqnarray}\label{eqvvv}
V \left (
\begin{array}{lcr}
0 &  & \\
& 0 & \\
 &  & {\delta_L}
 \end{array}
 \right )V^{\dagger}
 &=&\left (
 \begin{array}{lcr}
 s_{13}^2           & s_{13}s_{23}c_{13} & s_{13}c_{23}c_{13} \\
 s_{13}s_{23}c_{13} & s_{23}^2c_{13}^2 & s_{23}c_{13}^2c_{23} \\
 s_{13}c_{23}c_{13} & s_{23}c_{13}^2c_{23} & c_{23}^2c_{13}^2
 \end{array}
 \right )\delta_L \nonumber \\
 &=&\left (
 \begin{array}{lcr}
 0.0035 & -0.049 &0.032 \\
 -0.049 & 0.701  &-0.455 \\
 0.032  & -0.455 &0.296
 \end{array}
 \right )\delta_L ,
 \end{eqnarray}
where $\sqrt{m_{e}/m_{\mu}}\approx 0.0696$ and
 $m_{\mu}/m_{\tau}\approx 0.0594$ are used.
 Note that the corrections to the
 vertices of $Zee$, $Ze\mu$ and $Ze\tau$ are suppressed much more than
 the second and third family charged leptons,
 since the angle $\theta_{13}$ is approximated to zero
 compared with $\theta_{12}$ and $\theta_{23}$.
 The right hand part can be calculated in the same way.

%%%%%%%%%%%%%%%%%%%%%%%%%%%%%%%%%%%%%%%%%p14%%%%%%%%%%%%%%%%%%%%%%%%%%%%%%%%%
\section{Bounds on anomalous magnetic and electric moments {\normalsize \cite{Escribano,huang}}}
In general, the anomalous magnetic and the electric
dipole moments of the tau lepton are defined by (which we follow the
notation
of Ref.~\cite{Escribano})
\begin{eqnarray}\label{eq1}
a_\tau = \frac{g_\tau -2 }{2} = F_2(q^2=0), ~~{\rm and} ~~d_\tau = e {\tilde
F}_2(q^2 = 0),
\end{eqnarray}
where
$F_2$ and $\tilde F_2$ are form factors in the electromagnetic matrix element
\begin{eqnarray}\label{eq2}
< p^\prime | J_{em}^\mu(0) | p> = e {\bar u}(p^{\prime}) (F_1 \gamma^\mu +
[ \frac{i}{2 m_\tau} F_2 + \gamma_5 {\tilde F}_2]
\sigma^{\mu\nu} q_\nu ) u(p) ,
\end{eqnarray}
where $q = p^\prime - p $ and $ F_1(q^2=0 ) = 1$.

Theoretically the standard model predicts
$a_\tau = 1.1769(4) \times 10^{-3}$ and a very tiny $d_\tau$ from CP
violation in the quark sector~\cite{Czarnecki}.
%%%%%%%%%%%%%%%%%%%%%%%%%%%%%%%%%%%%table%%%%%%%%%%%%%%%%%%%%%%%%%
%\begin{center}
%\begin{table}[htb]
%\setlength{\tabcolsep}{1.0pc}
%\caption{Theoretical predictions and experimental measurements of the
%         anomalous magnetic and electric dipole moments of the electron
%        and muon.}
%\label{tab:eandmu}
%\begin{tabular*}{0.67\textwidth}{llc}  \hline
%
%$a_{\mathrm{e}}^{\mathrm{\scriptscriptstyle{SM}}}$ & $=0.001\,159\,652\,46(15)$  & \\
%$a_{\mathrm{e}}^{\mathrm{expt}}$                   & $=0.001\,159\,652\,193(10)$ & \\  \hline
%
%$a_\mu^{\mathrm{\scriptscriptstyle{SM}}}$          & $=0.001\,165\,920\,2(20)$   & \\
%$a_\mu^{\mathrm{expt}}$                            & $=0.001\,165\,923\,0(84)$   & \\ \hline
%
%$a_\tau^{\mathrm{\scriptscriptstyle{SM}}}$          & $=0.001\,177\,3(3)$   & \\
%$a_\tau^{\mathrm{expt}}$                            & $>-0.052~and~<0.058(CL=95\%)$& \\ \hline
%
%$d_{\mathrm{e}}^{\mathrm{expt}}$& $=(-2.7 \pm 8.3) \cdot 10^{-27}\,e\,{\mathrm{cm}}$  & \\  \hline
%
%$d_\mu^{\mathrm{expt}}$         & $=( 3.7 \pm 3.4) \cdot 10^{-19}\,e\,{\mathrm{cm}}$  & \\  \hline
%
%$d_\tau^{\mathrm{expt}}$         & $(>-3.1~and~<3.1)\times10^{-16}\,e\,{\mathrm{cm}}(CL=95\%)$  & \\
%\hline
%
%\end{tabular*}
%\end{table}
%\end{center}
%%%%%%%%%%%%%%%%%%%%%%%%%%%%%%%%%%%%%%%%%p15%%%%%%%%%%%%%%%%%%%%%%%%%%%%%%%%%

 The most
 stringent bounds are that inferred from the width $\Gamma ( Z \rightarrow
 \tau^+
 \tau^- )$, which are $-0.004 \leq a_\tau \leq 0.006$ and
 $| d_\tau | \leq 1.1 \times 10^{-17} {\it e}~{\rm cm}$~\cite{Escribano}.
 To obtain these limits, Escribano and Mass\'{o}
 took two effective operators ${\cal O}_{\tau B \phi}$ and ${\cal O}_{\tau W\phi}$
 in Eq.~(\ref{operator}).
 % approach to new physics,
 %
 %\begin{eqnarray}\label{eq3}
 %{\cal L}_{eff} = {\cal L}_0 + \frac{1}{\Lambda^2} \sum_i C_i {\cal O}_i ,
 %\end{eqnarray}
 %
 %where
 %${\cal L}_0$ is the SM Lagrangian, $\Lambda$ is the new physics
 %scale and $O_i$ are
 %$SU_c(3)\times SU_L(2)\times U_Y(1)$ invariant operators. $C_i$ are
 %constants which represent the coupling strengths of $O_i$.
 %Regard the anomalous magnetic moment of
 %the tau lepton,
 %
 %\begin{eqnarray}\label{eq4}
 %{\cal O}_{\tau B}&=&{\bar L}\sigma^{\mu\nu} \tau_R \Phi B_{\mu\nu}, \nonumber\\
 %{\cal O}_{\tau W}&=&{\bar L} \sigma^{\mu\nu} {\vec \sigma} \tau_R \Phi {\vec
 %W}_{\mu\nu},
 %\end{eqnarray}
 %
 %where $L=( \nu_{\tau}, ~ \tau_L )$, $\Phi$ is the Higgs scalar,
 %$B_{\mu\nu}$ and $W_{\mu\nu}$ are field strengths of $U_Y(1)$ and $SU_L(2)$.
 %Note that the operators involve only the
 %leptons of the third family and give rise to non-universal corrections to
 %interactions of the leptons.

 When $\Phi$ gets vacuum
 expectation value, operators ${\cal O}_{\tau B \phi}$ and
 ${\cal O}_{\tau W \phi}$ give rise to anomalous magnetic moment of the tau
 lepton and also corrections to the decay width of $Z$ into ${\bar \tau}
 \tau$. Given that the experimental data on Z width is quite consistent with the
 prediction of the SM, Escribano and Mass\'{o} put a strong bound on anomalous
 magnetic moment of the tau lepton listed above.

%%%%%%%%%%%%%%%%%%%%%%%%%%%%%%%%%%%%%%%%%p16%%%%%%%%%%%%%%%%%%%%%%%%%%%%%%%%%
 Considering the dimension-six operators,  ${\cal O}_{\tau B \phi}$ and
 ${\cal O}_{\tau W \phi}$, the full effective lagrangian now
 can be written as:

 \begin{eqnarray}\label{eq5}
 {\cal L}_{eff} = {\cal L}_0 + \frac{1}{\Lambda^2} ( c_{\tau B\phi} {\cal O}_{\tau B}
     + c_{\tau W \phi} {\cal O}_{\tau W} + h.c. ).
     \end{eqnarray}
     %
    % For the full listed effective operators in Eq.~(\ref{operator}, we find
     %that only ${\cal O}_{\tau B \phi}$ and
% ${\cal O}_{\tau W \phi}$ could contribute to the vertex $\gamma \tau \tau$ when
     After the electroweak symmetry is broken and the mass matrices of the fermions and the
     gauge bosons are diagonalized,
      the
      effective neutral current couplings of the leptons to gauge boson Z
      and the photon $\gamma$ are
\begin{eqnarray}\label{eq7}
{\cal L}^{Z, \gamma}_{eff}&=& eg^{Z, \gamma}
         {{ \pmatrix{{\overline e} \cr {\overline \mu} \cr
             {\overline \tau} \cr}}}^T  U_l \left \{
                 \pmatrix{ 1 & & \cr & 1 & \cr & &  1 \cr }
               ( \gamma_\mu V^{Z, \gamma} - \gamma_\mu \gamma_5 A^{Z, \gamma} )
               \right.   \nonumber \\
                 & &\left.  -\frac{1}{ 2 m_\tau}( i k_\nu \sigma^{\mu\nu}) S^{Z, \gamma}
                      \pmatrix{ 0 & & \cr
                         & 0 & \cr & &  1 \cr } \right \}U_l^\dagger
                          \pmatrix{e \cr \mu \cr \tau \cr}(Z^\mu,\gamma^\mu),
                           \end{eqnarray}
where $g^Z =  1/(4s_Wc_W), g^\gamma=1$, and
% In Eq.(\ref{eq7}),
$V^{Z,\gamma}=1-4s_W^2,1$,
$A^{Z,\gamma}=1,0$ for Z and photon respectively, and
%-------------------------------------------------------------------
\begin{eqnarray}
\label{eq32}
 S^Z&=&-\frac{8s_Wc_W}{e}\frac{m_\tau}{\Lambda^2}\frac{v}{\sqrt 2}\left [
    C_{\tau W \phi}c_W-2C_{\tau B\phi}s_W\right ],\nonumber\\
    \label{eq15}
     S^{\gamma}&=&\frac{2m_\tau}{e}\frac{\sqrt 2 v}{\Lambda^2}\left [
        C_{\tau W \phi}\frac{s_W}{2}-C_{\tau B\phi }c_W \right ].
    \end{eqnarray}

%%%%%%%%%%%%%%%%%%%%%%%%%%%%%%%%%%%%%%%%%p17%%%%%%%%%%%%%%%%%%%%%%%%%%%%%%%%%
The decay width of $l\rightarrow l'+\gamma$ is
given by
\begin{eqnarray}\label{eq500}
\Gamma_{(l\rightarrow l'\gamma)}=
\frac{m_l}{32\pi}\left(V_{ll'}e S^{\gamma} \frac{m_l}{m_{\tau}}\right)^2,
\end{eqnarray}
where $V_{ll^{\prime}}$ ($l\not=l^{\prime}$) are the nondiagonal
elements of matrix $V$ .
\begin{eqnarray}\label{eqvvv}
V = U_l \left (
\begin{array}{lcr}
0 &  & \\
& 0 & \\
 &  & 1
 \end{array}
 \right )U_l^{\dagger}
 \end{eqnarray}
We have neglected the
mass of the light lepton $l'$.

Given the current experimental upper limits on
$\mu^- \rightarrow e^-\gamma$, $4.9
\times 10^{-11}$ [13], we have
\begin{eqnarray}
\label{eq55}
\vert S^{\gamma} \vert < 1.3 \times 10^{-10}.
\end{eqnarray}
%

%%%%%%%%%%%%%%%%%%%%%%%%%%%%%%%%%%%%%%%%%p18%%%%%%%%%%%%%%%%%%%%%%%%%%%%%%%%%
The new physics contribution to the anomalous magnetic
moments of the tau lepton is given by
\begin{eqnarray}\label{eq17}
\vert \delta \alpha_\tau \vert =
\left \vert V_{\tau \tau} S^{\gamma} \right \vert.
\end{eqnarray}
With the bounds on $S^\gamma$ and $V_{\tau \tau}$, we obtain that
 \begin{eqnarray}
 \vert \delta \alpha_{\tau} \vert & \leq & 3.9 \times 10^{-11}.
 \end{eqnarray}
 This limit is much stronger than that
 that obtained by Escribano and Mass\'{o}. The limits from other
 LFV processes, such as $\tau^-\to e^- \gamma$, are weaker than that
given from $\mu^- \rightarrow e^-\gamma$.

Similarly, considering operators below which are introduced by Escribano and Mass\'{o},
 \begin{eqnarray}
 \label{eq23}
 \tilde{\cal O}_{\tau B\phi}&=&{\bar L}\sigma^{\mu\nu}i\gamma_5 \tau_R \Phi B_{\mu\nu}+h.c.,\nonumber \\
 \label{24}
 \tilde{\cal O}_{\tau W\phi}&=&{\bar L}\sigma^{\mu\nu}i\gamma_5{\vec \sigma} \tau_R \Phi
                    {\vec W}_{\mu\nu}+h.c.,
            \end{eqnarray}
and following the procedure above in obtaining the bounds on tau lepton magnetic
    moment, we put limit on the anomalous electric dipole moment
        \begin{eqnarray}
        \vert d_{\tau} \vert & \leq & 2.2 \times 10^{-25}~e~{\rm cm}.
        \end{eqnarray}
    Again this is stronger than that obtained by Escribano and Mass\'{o}.

Note the following relation:
\begin{eqnarray}\label{eq28}
{\vert \delta \alpha_\tau \vert}^2 \Gamma_{(\mu\to e\gamma)}
=\frac{8}{\alpha}\frac{m_\mu^3}{m_\tau^4}
\Gamma_{(\tau \to e \gamma)}\Gamma_{(\tau\to \mu\gamma)},
\end{eqnarray}
which is independent of the lepton mass mixing matrix~\cite{huang}.

%%%%%%%%%%%%%%%%%%%%%%%%%%%%%%%%%%%%%%%%%p19%%%%%%%%%%%%%%%%%%%%%%%%%%%%%%%%%

\section{ Effects on the decay $Z \to \mu^\pm + \tau^\mp$ }

The decays $Z\rightarrow \tau^+ + \tau^- $ and $Z\rightarrow \mu^+ + \mu^- $
can be discussed. For convenience, we define
\begin{eqnarray}\label{eq300}
M
\left(
\begin{array}{lcr}
0&0&0\\
0&0&0\\
0&0&1
\end{array}\right)M^{\dagger}
\equiv\left(
\begin{array}{lcr}
U_{ee}    &U_{e\mu}    &U_{e\tau} \\
U_{\mu e} &U_{\mu\mu}  &U_{\mu\tau}\\
U_{\tau e}&U_{\tau\mu} &U_{\tau\tau}
\end{array}\right),
\end{eqnarray}
where $M$ is $U$ or $V$.
Defining $\delta
\Gamma_{l \bar l}$ ($l= e,\mu,\tau$) to be the pure correction of the new
physics beyond the SM to the $Z\rightarrow l ~\bar l $ width,
$\Gamma_{l\bar l}$, we have
%----------------------------------------------------------------------
\begin{eqnarray}
\frac{\delta \Gamma_{l \bar l}} {\Gamma_{l \bar l}} \simeq
 \frac{2U_{ll}(g_L \delta_L +g_R \delta_R)}{g_L^2 + g_R^2}
 = (-4.28\delta_L+3.68\delta_R)U_{ll},
 \end{eqnarray}
 %----------------------------------------------------------------------
 where $U_{ll}$ ($l=e,\mu,\tau$) are diagonal elements of matrix
, and we have used $\sin^2\theta_W=0.231$.
 It is clear that different
 corrections of width would be given by using different approximations.

 The constraints of the parameters $\delta_L$ and $\delta_R$ can be given
 in the following two cases in term of the
 experimental data of the decays,
 $$
 \Gamma_{\mu \bar
 \mu}/\Gamma_{full}=(3.367\pm 0.013)\%,
 $$
 $$\Gamma_{\tau \bar
 \tau}/\Gamma_{full}=(3.360\pm 0.015)\%.
 $$
%%%%%%%%%%%%%%%%%%%%%%%%%%%%%%%%%%%%%%%%%p20%%%%%%%%%%%%%%%%%%%%%%%%%%%%%%%%%

The width of the FCNC decays can be given by
\begin{eqnarray}
\Gamma_{ll^{\prime}} = \frac{\vert U_{ll^{\prime}} \vert^2}{6\pi v^2}
        m_Z^3(\delta_L^2+\delta_R^2),
    \end{eqnarray}
where $U_{ll^{\prime}}$ ($l\not=l^{\prime}$) are the nondiagonal elements,
and the mass of charge lepton is approximated to zero compared with the
heavy mass of $Z$.

Under the constraints derived from
$\frac{\delta \Gamma_{l \bar l}} {\Gamma_{l \bar l}}$, we can get the
limit of the FCNC effects.

According to the our calculation, it is found that the limit of FCNC is
directly related to the experimental data of the decays
$Z\rightarrow \tau^+ + \tau^- $ and $Z\rightarrow \mu^+ + \mu^- $
. Under assumption that the experimental error of the width
$Z\rightarrow \tau^+ + \tau^- $ and $Z\rightarrow \mu^+ + \mu^- $
is contributed
mainly by the effects of new physics , the branching ratio of the decay
$Z\rightarrow \mu^\pm + \tau^\mp $ could reach to $10^{-6}-10^{-7}$.

This conclusion is larger than that from the calculation of
loop level in the SM, which could be $10^{-7}-10^{-8}$.

Since the recent experimental
data of the branching ratio of the rare decay $Z\rightarrow \mu^\pm +
\tau^\mp$ is less than $1.2\times 10^{-5}$, our prediction can
be experimentally accessible and hopeful for observation in the future colliders.

%%%%%%%%%%%%%%%%%%%%%%%%%%%%%%%%%%%%%%%%%p20%%%%%%%%%%%%%%%%%%%%%%%%%%%%%%%%%

\section{Summary}

\begin{enumerate}
\item The phenomena of the lepton FCNC in $Z$ decays by means of
the effective lagrangian and the lepton flavor-mixing matrices are studied.
The dimension-six CP-conserving $SU_L(2)\times
U_Y(1)$ invariant operators involving the tau lepton,
generated by new physics at a higher energy scale, are listed.
The phenomena of the lepton FCNC and the lepton number violation are subject to
the existing experimental limits and are hopeful for observation in the
future collider.

\item  We extend the work by Escribano and Mass\'{o} to bound the anomalous magnetic
and
electric dipole moments of the tau lepton in the effective lagrangian by allowing the
mixing of three generations in the lepton sector. In
 the standard model, these mixing effects
 are not measurable because of
 vanishing neutrino masses and the universal
 gauge interactions. With
 non-universal interaction, the lepton flavor violation
  happens even with zero neutrino
  masses.

  By taking the lepton mass matrix of Fritzsch ansatz,
   we have demonstrated that the
   experimental limit on $\mu \rightarrow e \gamma$ puts stronger
    limits on the anomalous
    magnetic and electric dipole moments of the tau lepton than obtained
     by Escribano and Mass\'{o}.

 \begin{eqnarray}
 \vert \delta \alpha_{\tau} \vert  \leq  3.9 \times 10^{-11},~~~
  \vert d_{\tau} \vert  \leq  2.2 \times 10^{-25}~e~{\rm cm}.
  \end{eqnarray}

\item Our results depend on the
lepton mass matrix. So qualitatively our results also indicates the
equal imporance of probing the anomalous magnetic
and electric dipole moments of tau lepton
as well as the lepton flavor violation. Furthermore,
the future experimental data on
anomalous magnetic and electric dipole moments of the tau lepton together with
lepton flavor violation will provide an experimental test on various
lepton mass ansatz.

\item Among all of 10 such operators, 6 operators ($O^{(1)}_{\Phi
L}$, $O^{(3)}_{\Phi L}$, $O_{LB}$, $O_{L W}$, $O_{\Phi \tau}$
and $O_{\tau B}$) contribute mainly to the $Z\tau\tau$ coupling.
Their corresponding constants $C^{(1)}_{\Phi L}$, $C^{(3)}_{\Phi L}$,
$C_{LB}$, $C_{L W}$, $C_{\Phi \tau}$ and $C_{\tau B}$ are contained by two
parameters $\delta_L$ and $\delta_R$, which could be constrained by the
experimental data of the decays
$Z\rightarrow \tau^+ + \tau^- $ and $Z\rightarrow \mu^+ + \mu^- $.
According to these constraints, the branching ratio of the decay
$Z\rightarrow \mu^\pm + \tau^\mp $ could reach to $10^{-6}-10^{-7}$, which
is 100 times larger than that of $Z\rightarrow e^\pm + \tau^\mp $ and
$Z\rightarrow e^\pm + \mu^\mp $.

\end{enumerate}

%=======rrr===================================================================

\eject
%%%%%%%%%%%%%%%%%%%%%%%%%%%%%%%%%%%%%%%%%%%%%%%%%%%%%%%%%%%%%%%%%%%%%%%%%

\begin{thebibliography}{99}
\bibitem{Hewett}
%[1]
            R.D. Peccei, hep-ph/9811309; J. L. Hewett, hep-ph/9810316
\bibitem{Whisnant}
%[2]
                K. Whisnant, J. M. Yang, B.-L. Young and X. Zhang,
                 Phys. Rev. D {\bf 56}, 467 (1997).
\bibitem{Yang}
%[3]
        J. M. Yang and  B.-L. Young,
                  Phys. Rev. D {\bf 56}, 5907 (1997).
\bibitem{Zhang}
%[4]
        X. Zhang and B.-L Young,
                 Phys. Rev. D {\bf 51}, 467 (1995)
\bibitem{Burgess}
%[4+1]
                C. J. C. Burgess and H. J. Schnitzer,
                 Nucl. Phys. B {\bf 228}, 454 (1983);
                 C. N. Leung, S. T. Love and S. Rao, Z. Phys. C {\bf 31},
                 433 (1986);
                 W. Buchmuller and D. Wyler, Nucl. Phys. B {\bf 268},
                 621 (1986).
\bibitem{Young}
%[5]
                T. Huang, J. M. Yang, B.-L. Young and X. Zhang,
                Phys. Rev. D {\bf 58}, (1998) 073007.
%\bibitem{Totsuka}
%[6]
%                 Y. Totsuka, talk given at the 18th International
 %               Symposium on Lepton-Photon Interactions; Hamburg,
 %               July 1997' E. Kearns, talk given at the ITP Conference on
 %               Solar Neutrinos:  {\sl News About SNUs},
 %               Santa Barbara, December 1997;
 %               Super-Kamiokande Collaboration, hep-ex/9807003.
\bibitem{Fritzsch}
%[7]
         H. Fritzsch, Nucl. Phys. B {\bf 155} (1979) 189.
\bibitem{Xing1}
%[8]
         H. Fritzsch and Z. Z. Xing,
                 Phys. Lett. B {\bf  372} (1996) 265.
\bibitem{Xing2}
%[9]
         H. Fritzsch and J. Plankl,
         Phys. Lett. B {\bf 237} (1990) 451;
         H. Fritzsch and D .Holtmannsp\"{o}tter,
         Phys. Lett. B {\bf 338} (1994) 290;
         H. Fritzsch and Z. Z. Xing, hep-ph/9808272,
         to be published in Phys. Lett. B;
                 H. Fritzsch and Z. Z. Xing, hep-ph/9807234;
         M. Fukugita, M. Tanimoto and T. Yanagida,
         Phys. Rev. D {\bf 57}, (1998) 4429;
         M. Tanimoto, hep-ph/9807283.
\bibitem{Xing3}
%[10]
        For example, Z. Xing, hep-ph/9804433.
\bibitem{Escribano}
%[13]
        R. Escribano and E. Masso,
                 Phys. Lett. B{\bf 395}, 369 (1997).
\bibitem{huang} T. Huang, Z.-H. Lin and X. Zhang, Phys. Lett. B{\bf 450}, 257 (1999).

\bibitem{Czarnecki}
%[12]
        A. Czarnecki and W. J. Marciano,
                  hep-ph/9810512.
\bibitem{Caso}
%[11]
        C.Caso et al., Particle Data Group, Eur. Phys. J.
          {\bf C 3} (1998) 1.
%\bibitem{Taylor}
%[14]
%        L. Taylor, invited talk at the TAU'98
%               Workshop, 14-17 September 1998, Santander,
%               Spain, hep-ph/9810463.
%\bibitem{Pham}
%[15]
%                X.-Y. Pham, hep-ph/9809322
\end{thebibliography}
\end{document}